\documentclass[12pt]{article}
\usepackage{amssymb}
\usepackage{amsfonts}
\usepackage{graphicx}

\def\beq{\begin{equation}}
\def\eeq{\end{equation}}
\def\bb{\begin{eqnarray}}
\def\ee{\end{eqnarray}}
\def\dz{\partial_z}

\begin{document}
\begin{titlepage}

\begin{center}
{\Large\bf Vacua Landscape Attractor}
\end{center}
\vspace{0.5cm}
\begin{center}

{\large Alvaro N\'u\~nez\footnote{e-mail address:
an313@scires.nyu.edu}  and Slava Solganik\footnote{e-mail address:
ss706@scires.nyu.edu}}
\vspace{0.5cm}

{\em New York University, Department of Physics, New York, NY 10003, USA}.\\
\end{center}

\hspace{1.0cm}

\begin{abstract}
\noindent The recent progress in the understanding of the landscape of string theory vacua
hints that the hierarchy problem might be the problem of a super-selection rule.
The attractor mechanism  gives a possibility to explain the choice  of a vacuum.
We consider a toy model of self-interacting membranes and show that
for a very generic interaction there are attractor solutions.
\end{abstract}

\end{titlepage} 

\newpage 

\bigskip 

\section{Introduction}

The hierarchy problem manifests itself in the enormous difference between 
the standard model, gravity and in a wider sense the dark energy scales.
It is assumed often that in the case of the standard model its solution requires some UV regulating physics. 
However, as it was suggested in~\cite{Dvali:2003br}-\cite{Dvali:2004tm}, the hierarchy problem can be addressed as the problem of 
a vacuum super-selection rule. The recent progress in the understanding of the string theory vacua 
landscape (see for example~\cite{Giddings:2001yu}) gives a hint on the different possibilities 
of vacua density distributions. This motivates the studying of an alternative mechanism as a possible solution to the hierarchy problem.
The idea proposed in~\cite{Dvali:2004tm}, based on an earlier work on
cosmic attractors~\cite{Dvali:2003br}, is to put to work the multiplicity of vacua.
The hierarchy problem  is promoted into a problem of the super-selection
rule among the infinite number of vacua, that are finely scanned by the Higgs mass. In this
framework, the Higgs mass is promoted into a dynamical variable. 
An infinite number of vacua cluster around a certain point making it an attractor.
On the resulting landscape in all but a measure zero set of
vacua the Higgs mass has a common, hierarchically small  value due to the attractor.

In this paper we will  analyze a model which can be viewed as an effective theory obtained after integrating out the Higgs field in the model~\cite{Dvali:2004tm}.
As a result we get a setup in which branes adjust their charges according to the values of the field they produce.
We consider the possibility of having an attractor for such system.
Three-forms are sourced out by a number of membranes (two-branes) with charges $q$ that can be
self-adjusted. In particular, we will consider the membrane charges being tuned
as an arbitrary function of the field. This problem essentially can be
reduced to that of field dependent charges in 1+1 electrodynamics. As we will show,
the presence of an attractor is a very generic feature of such models.

\section{Three-forms and two-branes}
Let's review the setup of the work~\cite{Dvali:2004tm}. 
The spectrum of different string theories contains antisymmetric form fields,
which after compactification to four dimensions give rise to three-forms, two-forms and one-forms. 
In particular, we are interested in three-forms $C_{\alpha\beta\gamma}$.
The action for a three-form in four dimensions reads
\beq
S_C=\int_{3+1} \, {1 \over 48} \, F_{\mu\alpha\beta\gamma}F^{\mu\alpha\beta\gamma},
\label{F_action}
\eeq
where the four-form field strength
\beq
F_{\mu\alpha\beta\gamma}\, = \, d_{[\mu}C_{\alpha\beta\gamma]}.
\eeq 
This action is gauge invariant, and this guarantees the decoupling of the time component. The gauge transformation is
\beq
C_{\alpha\beta\gamma}  \rightarrow  C_{\alpha\beta\gamma} \, + \, d_{[\alpha}\Omega_{\beta\gamma]},
\eeq
where $\Omega_{\beta\gamma}$ is some two-form depending on the coordinates and the square brackets denote anti-symmetrization. 
The form $C$ has no propagating degrees of freedom in four dimensions. The equations of motion stemming from the action~(\ref{F_action}) are
\beq
\partial^{\mu}F_{\mu\nu\alpha\beta} \, = \, 0 
\eeq
and have a constant solution
\beq
F_{\mu\nu\alpha\beta} \, = \, F_0 \,  \epsilon_{\mu\nu\alpha\beta};
\eeq
here  $F_0={\rm constant}$ and $\epsilon_{\mu\nu\alpha\beta}$ is the totally antisymmetric tensor. In the absence of interactions with other fields
this constant changes the Lagrangian and contributes to the cosmological term. In the presence of interactions $F_0$ will contribute to those fields masses
and to the couplings.

Three-forms couple to two-branes, e.g. membranes. The effective action is given by
\beq
S={q\over 6}\, \int_{2+1}  d^3\xi \, C_{\mu\nu\alpha} \left( {\partial Y^{\mu} \over \partial \xi^a}
{\partial Y^{\nu} \over \partial \xi^b}{\partial Y^{\alpha} \over \partial \xi^c}\,\right) \epsilon^{abc}
   \, - \,  \int_{3+1} \, {1 \over 48} F^2.
\eeq
where $q$ is the brane charge and $Y^\mu(\xi)$ describe the brane history as a function of its world volume coordinates $\xi^a,\quad a=0,1,2$
We can rewrite the interaction term as a four dimensional integral
\beq
\int \, d^4x \,  {1 \over 6} J^{\alpha\beta\gamma} C_{\alpha\beta\gamma}
\eeq
where the brane current
\beq
J^{\alpha\beta\gamma}(x)\,  = \,  \int d^3\xi \delta^4(x \, - \, Y(\xi))\, q \,  
 \left( {\partial Y^{\alpha} \over \partial \xi^a}
{\partial Y^{\beta} \over \partial \xi^b}{\partial Y^{\gamma} \over \partial \xi^c}\,\right) \epsilon^{abc}.
\eeq 
As long as $q$ is constant, the current is conserved. We end up with the equations of motion 
\beq
\partial_{\mu}\,  F^{\mu\nu\alpha\beta} \, = \, -\, q\int d^3\xi \delta^4(x \, - \, Y(\xi)) 
 \left( {\partial Y^{\nu} \over \partial \xi^a}
{\partial Y^{\alpha} \over \partial \xi^b}{\partial Y^{\beta} \over \partial \xi^c}\,\right) \epsilon^{abc}.
\eeq
We consider the simple case of static and flat branes,
\beq
\;\quad\qquad\qquad Y^{\mu} \, = \, \xi^{\mu},\qquad\mu = 0,1,2
\eeq 
\beq
Y^3 \, = \, 0\qquad
\eeq
We take $x_3=z$ as the coordinate transversal to the brane. Then the equations of motion reduce to
\beq
\partial_{\mu}\,  F^{\mu\nu\alpha\beta} \, = \, -\, q \delta(z) \epsilon^{\nu\alpha\beta z}.
\eeq 
The equations of motion show that the brane separates two vacua. In each of them the field strength is constant and the 
jump between the values of the field in different vacua is given by the brane charge $q$. This way, there is a solution with multiplicity of vacua and the vacua in this solution are labeled by an integer $n$,
\beq
-{1\over 24}F_{\alpha\beta\gamma\mu}\epsilon^{\alpha\beta\gamma\mu} \, = \, qn \, + \, F_0,
\eeq
where $F_0$ is a constant which in the theory with an attractor mechanism will be fixed. 

In the model~\cite{Dvali:2004tm}, the lowest order parity and gauge-invariant Lagrangian describing a non-trivial interaction 
between the Higgs field $\phi$ and the gauge field $C$ was suggested in the form
\begin{equation}
\label{fphi}
L \, = \, |\partial_{\mu}\phi|^2 \,  - \,  {1 \over 48} F^2 \, + \,  \, |\phi|^2 \, \left (m^2\,  + \,  {F^2 \over 48M^2}
\right) \, - \, {\lambda \over 2} \, |\phi|^4
\, + \, ...
\end{equation}
where $\lambda$ is the quartic coupling and $m$,  $M$ are mass parameters. 
As a result, the gauge field determines the value of the effective mass and 
consequently the  vacuum expectation value of the Higgs field.
The Higgs field in turn readjusts the brane charges and gauge field  closing a cycle.
This can create an attractor depending on details of the interaction provided there is an 
additional symmetry forbidding higher loop corrections to the classical attractor.

\section{Explicit gauge field dependence}
The key idea is to consider a charge $q(F)$ being explicitly field dependent; 
this corresponds to effective ``integrating out'' the Higgs mass in the model~\cite{Dvali:2004tm}.
To simplify the derivations and make the physical content clearer, we will consider the 1+1 case and
will call the field potential $A$. 
Thus we will have electrodynamics with self-adjusting charges. 
In the absence of the mentioned dependence the current is
\beq
J^{\mu}(x)=\int d\xi\,q\,\delta^2(x-Y(\xi))\frac{\partial Y^\mu(\xi)}{\partial\xi},
\eeq
where the charge $q$ acts as source for the gauge field. 
If we consider the charges being field depending sources, the field in turn readjusts the charge. 
The corresponding current can be written as
\beq
J^{\mu}(x)=\int d\xi\,q\left(F(Y(\xi))\right)\,\delta^2(x-Y(\xi))\frac{\partial Y^\mu(\xi)}{\partial\xi}.
\eeq
This current is no longer conserved unless we rewrite the interaction so that the field
couples only to the transverse part of the current. This can be guaranteed by an interaction term with a projection kernel
\beq
\Pi_{\mu\nu}=g_{\mu\nu}-\frac{\partial_\mu\partial_\nu}{\partial^2}
\eeq
so that the Lagrangian
\beq
{\cal L}=-\frac14F_{\mu\nu}F^{\mu\nu}+A^\mu\Pi_{\mu\nu}J^\nu.
\eeq
The potential $A_\mu$ couples only to the transverse part of the current $J^{\mu}$. For a single static charge current located at point $x_1\equiv z=a$
\beq
J^0(z)=q(F(z))\delta(z-a),
\quad J^1(z)=0
\eeq
The interaction term
\beq
J^0\Pi_{0\nu}A^\nu=J^0\Pi_{00}A^0=J^0A_0-J^0\frac{\partial_0\partial_0}{\partial^2}A_0=J^0A_0
\eeq
and the Lagrangian becomes
\beq
{\cal L}=-\frac12(\partial_z A_0)^2+q(\partial_z A_0)\delta(z-a)A_0.
\eeq
The variation of the Lagrangian is
\beq
\delta{\cal L}=(\partial_z^2 A_0+q(\partial_z A_0)\delta(z-a)-\dz(q'(dz A_0)\delta(z-a)A_0))\delta A_0
\eeq
and gives the equation of motion
\bb
\dz^2 A_0=\delta(z-a)\left[\dz A_0q'(\dz A_0)-q(\dz A_0)+q''(\dz A_0)\dz^2A_0A_0\right]+\nonumber \\
+\dz\delta(z-a)q'(\dz A_0)A_0.
\ee
Outside the brane the equation of motion reduces to
\beq
\dz^2A_0=0.
\eeq
The general solution for the field strength is a constant,
\beq
F\equiv F_{10}=\dz A_0={\rm const}.
\eeq
Integrating in a small neighborhood near the brane leads to the boundary condition
\beq
F(a+0)-F(a-0)=\qquad\qquad\qquad\qquad\qquad\qquad\qquad\qquad\qquad
\eeq
\beq
=\big[F q'(F)-q(F)+q''(F)\dz F A_0-\dz(q'(\dz A_0)A_0)\big]_{z=a}
\eeq
or equivalently 
\beq
F(a+0)-F(a-0)=-q(F)\big|_{z=a}.
\label{bc}
\eeq
We can implement this boundary condition into the equation of motion as 
\beq
\dz F=-q(F(a))\delta(z-a)
\eeq
Let's look at the behavior of the field when we add in succession $N$ charged branes at 
points $z=a_n$ (for convenience we take $a_n>a_k$ for $n>k$). 
In this case we get the following equation
\beq
\dz F=-\sum_{k=1}^N q(F(a_k))\delta(z-a_k)
\eeq
where the integer $k=1,\dots, N$ labels the branes. The general solution to this equation is
\beq
F(z)=-\sum_{k=1}^N q(F(a_k))\theta(z-a_k)
\eeq
with the boundary condition to be applied. These boundary conditions will fix the constants $q(F(a_k))$. 
The equations of motion lead to the following recursion relation
\beq
F_k+\frac12q(F_k)=F_{k-1}-\frac12q(F_{k-1})
\label{recursion}
\eeq
\begin{figure}[t]
\begin{center}
\includegraphics{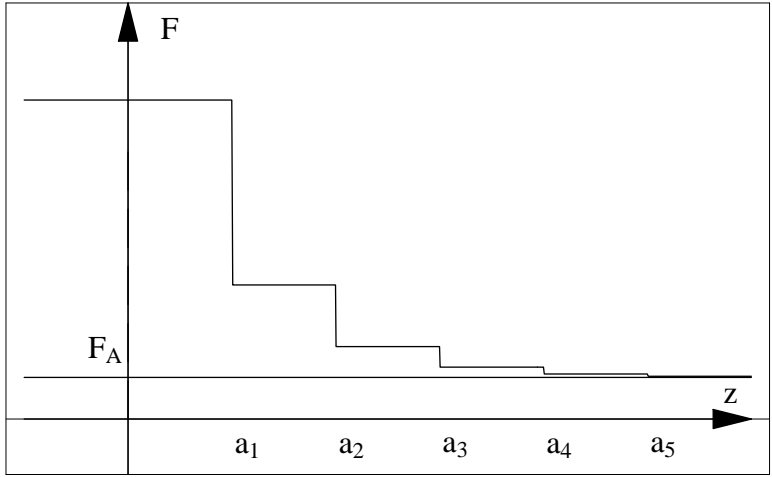}
\end{center}
\caption{{Field configuration with field-dependent charges, $0<c<2$.}}
\label{charges1}
\end{figure}
\begin{figure}[t]
\begin{center}
\includegraphics{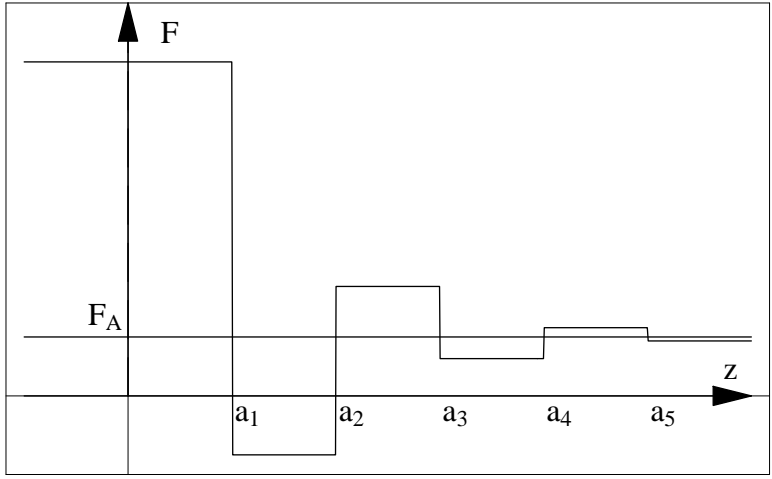}
\end{center}
\caption{{Field configuration with field dependent charges, $c>2$.}}
\label{charges2}
\end{figure}
for the values of the fields between the branes. As $N\to\infty$, if there is an attractor point, 
the following limit should exist 
\beq
\lim_{N\to\infty} F_N \to F_A .
\eeq
From the recursion one can see that when the limit exists, then $q(F_A)=0$. It means that the attractor point candidate $F_A$ should be a root  of the equation
\beq
q(F)=0.
\eeq
In Figures \ref{charges1} and \ref{charges2} we show how the addition of branes at points $a_1$, $a_2$,... leads to an increasing number of vacua near the attractor point $F=F_A$. A sufficient condition for the existence of the above limit, i.e. the attractor point at the root $F_A$, is $q'(F_A)>0$.

\section{Number Density near the Attractor Point}

We will evaluate the vacuum number density near the attractor point.
In the case of a charge depending linearly on the field $F$,  
\beq
q(F)=c(F-F_A)
\eeq 
with $c$ some positive constant, 
the recursion (\ref{recursion}) with an initial value $F=F_1$ has the solution
\beq
F_k=\left(\frac{2-c}{2+c}\right)^{k-1}\left(F_1-F_A\right)+F_A.
\eeq
We can express the number of vacuum states outside the interval of fields $F-F_A$ 
\beq
k=1+\frac{\ln\frac{F-F_A}{F_1-F_A}}{\ln\frac{2-c}{2+c}}.
\eeq
Correspondingly, the number density of the vacuum states in the linear case is given by
\beq
n\sim -\frac{d k}{dF}=-\frac1{\ln\frac{2-c}{2+c}}\frac1{|F-F_A|}
\eeq
and is divergent at the attractor point $F=F_A$.

\begin{figure}[t]
\begin{center}
\includegraphics{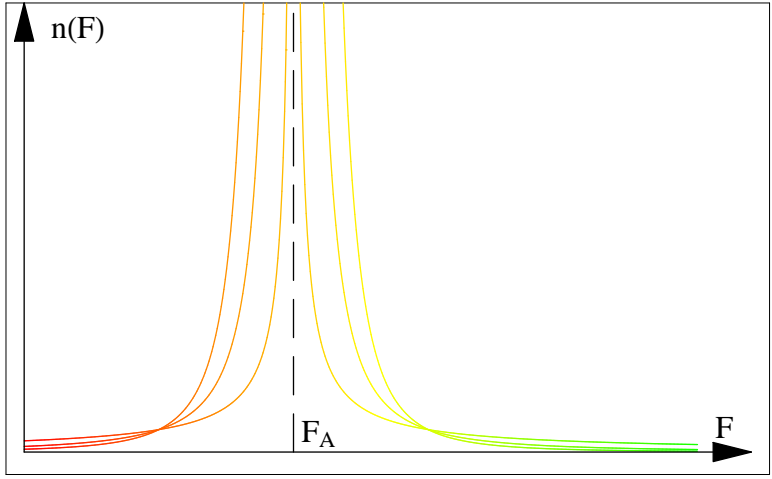}
\end{center}
\caption{The number of states density around the attractor point.}
\label{vacua}
\end{figure}

We would like to calculate the number density for arbitrary self-interaction but 
we cannot explicitly write the number of states $k$ in terms of the field range for a generic function $q(F)$. 
Nevertheless we can estimate the derivative via the recursion relation
\beq
n\sim\frac{\Delta k}{\Delta F}=\frac1{F_k-F_{k-1}}\sim\frac1{q(F)}
\eeq
This number of states is also divergent at the attractor point. 
In Figure (\ref{vacua}) we have depicted the typical, divergent 
behavior of the number density of vacua near the attractor point. 

Let's find the domain of convergence to the attractor. The answer comes from the sufficient condition of the attractor existence
at the point $F=F_A$ 
\beq
q(F_A)=0,\qquad q'(F_A)>0.
\eeq
It follows from the recursion relation that in the vicinity of the attractor point $F_A$ to guarantee the convergence we should continuously satisfy
\beq
q'(F)>0,
\eeq 
\beq
q'(F)<4+q'(F_A).
\eeq

In the linear example
\beq
q(F)=c(F-F_A)
\eeq
the fields converge everywhere for $c>0$.

For a quadratic dependence
\beq
q(F)=(F-F_A)(F-F_B)
\eeq
the attractor will be located at $F_A$ for
\beq
F_A>F_B.
\eeq 
The range of the fields $F$ which converge to the attractor point is defined by
\beq
0<q'(F)<4+q'(F_A).
\eeq 
This restricts $F$ to the range
\beq
(F_A+F_B)/2<F<2+F_A.
\eeq

\section{Conclusions}
 
We have shown that the self-adjusting charges or membranes have an attractor point. 
This implies that the number density of vacua within a small range around the attractor point  
blows up. From the physical point of view, the attractor adjusts to the point with a minimal 
self-interaction, creating an enormous number of vacua with close values. 
The sufficient condition for the attractor existence is vanishing of the charge at the 
attractor point and the charge being an increasing function of the field. 
Even if the interaction never reaches zero, this point still will be like an attractor. 
However, the number density of vacua will have a finite maximum sharp pick.

We hope that the suggested attractor model can be implemented for a natural explanation
of large hierarchies, like scales of electroweak theory, gravity and dark energy. \\

Acknowledgments:
it is pleasure to thank Gia Dvali for raising this problem and  for useful discussions.


\end{document}